\documentclass[doublecol]{epl2} 

\usepackage[utf8]{inputenc}
\usepackage{epsf,graphicx}
\usepackage{multirow}
\usepackage{amsmath,gensymb,amssymb}
\usepackage[bottom]{footmisc}
\usepackage{lipsum}
\usepackage{dcolumn}
\usepackage{bm}

\usepackage{hyperref}
\hypersetup{
colorlinks=true, 
citecolor=magenta,
breaklinks=true, 
urlcolor= blue, 
linkcolor= blue, 
bookmarksopen=true, 
}

\usepackage{color}
\definecolor{linkcolor}{rgb}{0,0,0.6}

\title{Experimental quasi-1D capillary-wave turbulence}
\author{Guillaume Ricard \and Eric Falcon\thanks{E-mail: \email{eric.falcon@u-paris.fr} (corresponding author)}}
\shortauthor{Guillaume Ricard and Eric Falcon}
\institute{
Universit\'e de Paris, MSC, UMR 7057 CNRS, F-75 013 Paris, France
}

\abstract{\ Wave turbulence in quasi-1D geometry is usually not investigated experimentally since low-order resonant wave interactions are theoretically prohibited. Here, we report on the first observation of unidirectional capillary-wave turbulence on the surface of a fluid in a canal. We also show that five-wave interactions are the lowest-order resonant process subsisting at small scales, and are thus probably the one generating such quasi-1D capillary wave turbulence. We show that the wave spectrum is compatible with the corresponding dimensional analysis prediction. The main assumptions of weak turbulence theory are also verified experimentally. Quasi-1D wave turbulence could be thus highlighted in other fields of wave turbulence.}

\begin{document}

\maketitle


\section{Introduction}
Wave turbulence is a phenomenon occurring within a large number of random nonlinear interacting waves~\cite{ZakharovBook1992,NazarenkoBook2011,NewellARFM2011}. These nonlinear interactions lead to an energy cascade from a large (forcing) scale down to a small (dissipative) scale, predicted by weak turbulence theory (WTT). These predictions have been applied in many different domains such as ocean surface waves, plasma waves, hydroelastic or elastic waves, internal waves, and optical waves~\cite{ZakharovBook1992,NazarenkoBook2011,NewellARFM2011}. 
This theory has then been assessed experimentally in various wave systems propagating in 2D or 3D
\cite{FalconDCDSB2010,ShriraBook,Hawai,YaromNat2014,MonsalvePRL2020,SavaroPRF2020}, but rarely in 1D
since generally no low-order resonant wave interactions are expected theoretically in this geometry~\cite{NazarenkoBook2011}. To our knowledge, the unique experimental study concerns 1D nonlinear optics focusing only on inverse cascade towards large scales~\cite{BortolozzoJOSA2009}. For capillary waves on the surface of a fluid, the WTT \cite{ZakharovJAMTP1967,GaltierGAFD2020} is rather well confirmed experimentally in 2D (see review~\cite{FalconARFM2022}), whereas for a unidirectional propagation, the theory forbids low-order wave resonant interactions~\cite{NazarenkoBook2011,ZakharovPR2004}, and thus a wave turbulence regime. Nevertheless, a quasi-1D capillary wave turbulence regime has been recently reported numerically~\cite{KochurinJETP2020}. 
An experimental observation of such a regime would pave the way to other fields of wave turbulence due to easier calculations, and measurements in 1D geometry.  

In this letter we report the first observation of quasi-1D capillary-wave turbulence on the surface of a low-viscous fluid (mercury). With this specific fluid, a weak nonlinearity is sufficient to reach a wave turbulence regime without coherent structures. Using high-order correlations of wave elevations, we quantify the occurrence of three-, four-, and five-wave interactions. Although {\em quasi-resonant} interactions are observed at low orders, five-wave {\em resonant} interactions are found to be the lowest-resonant order subsisting in the capillary range, and are thus probably the mechanism generating the observed 1D capillary-wave turbulence. This differs from the usual 2D capillary-wave turbulence involving three-wave resonant interactions \cite{NazarenkoBook2011,ZakharovJAMTP1967}. 
We also show that the wave spectra in frequency and in wavenumber are compatible with the corresponding dimensional analysis predictions. Moreover, the energy flux cascading towards small scales is roughly found to be constant as expected, and the main WTT assumptions are verified experimentally. Note that this 1D wave turbulence differs basically from 1D integrable turbulence (involving coherent structures such as solitons within stochastic waves)~\cite{Zakharov1971}, recently observed~\cite{CazaubielPRF2018}.  
Note also that an idealized 1D model of wave turbulence showed strong deviations from the WTT due to these coherent structures \cite{ZakharovPR2004,MMT1997}. Moreover, our results should not be confused with the existence of unidirectional resonant interaction highlighted in 2D gravity-capillary wave turbulence, near the gravity-capillary crossover~\cite{AubourgPRF2016}. 

 
 

\section{Theoretical backgrounds}
The dispersion relation of linear deep-water gravity-capillary surface waves reads $\omega^2=gk+(\gamma/\rho)k^3$, with $\omega=2\pi f$ the angular frequency, $k$ the wave number, $g$ the acceleration of gravity, $\gamma$ the surface tension, and $\rho$ the liquid density~\cite{Lamb1932}. The theoretical crossover between the gravity and capillary regimes occurs for $k_{gc}=\sqrt{\rho g/\gamma}$ and $f_{gc}=(g^3\rho/\gamma)^{1/4}/(\sqrt{2}\pi)$~\cite{FalconPRL2007}. Both contributions coexist near $f_{gc}$ \cite{FalconARFM2022}, and we denote $f_c$ the frequency from which gravity becomes negligible (see below). In the two limits of weak nonlinearity and infinite system, WTT predicts a wave spectrum cascading from large to small scales. The energy transfers between waves occur due to a $N$-wave resonant interaction process, $N$ being fixed by the geometry and the wave system considered~\cite{NazarenkoBook2011}. For 1D gravity waves, WTT predicts that $N=5$~\cite{dyachenko1995five}. 
For 1D capillary waves, no prediction exists up to now. In 1D, a $N$-wave \textit{resonant} interaction process requires to satisfy

\begin{equation}
k_1\pm k_2 \pm ...\pm k_N=0{\rm , \ and \ \ } \omega_1\pm \omega_2\pm ...\pm \omega_N=0 {\rm ,}
\label{pureIntN4}
\end{equation}
where $k_i$ and $\omega_i$ take positive values, $N\geq 3$, and $\omega_i \equiv \omega(k_i)$. No nontrivial solution of Eq.~(\ref{pureIntN4}) exists with $N=3$ or $4$ for 1D capillary waves, i.e., for $\omega(k_i)=\sqrt{\gamma/\rho}k_i^{3/2}$. 
Nonlinearity can broaden the dispersion relation authorizing other possible interactions (called \textit{quasi-resonant} interactions) as

\begin{equation}
k_1\pm k_2 \pm ...\pm k_N=0{\rm , \ and \ \ } \omega_1\pm \omega_2\pm ...\pm \omega_N<\delta_\omega{\rm ,}
    \label{quasiIntN4}
\end{equation}
where $\delta_\omega$ is a constant corresponding to the nonlinear wave frequency broadening. It is important to emphasize that WTT needs resonant interactions to build wave turbulence: even if quasi-resonant interactions occur at some specific order, the nonlinear process generating wave turbulence corresponds to the lowest nonlinear order for which resonant interactions exist, as a consequence of the change of canonical variables \cite{ZakharovBook1992,NazarenkoBook2011}. In the particular case where resonant interactions are absent (e.g., by numerically truncating to a specific nonlinear order), quasi-resonances can generate an energy cascade \cite{KochurinJETP2020}. 

\begin{figure}[t!]
    \centering
    \includegraphics[width=1\linewidth]{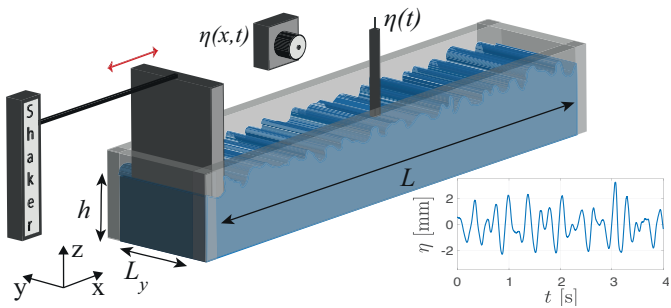} 
    \caption{Experimental setup. The wave elevation, $\eta(t)$, is measured at a single point using a capacitive wire gauge, and resolved in space and time, $\eta(x,t)$, with a lateral camera. Inset: typical wave elevation signal $\eta(t)$ for $\epsilon=0.06$.}
    \label{setup}
\end{figure}

Using dimensional analysis~\cite{Connaughton2003,KochurinJETP2020}, one can predict the power spectra of the wave elevation for capillary waves for a $N$-wave process as
\begin{equation}
        \begin{split}
        &S_{\eta}(\omega) =C_{KZ}^\omega \langle P \rangle^{\frac{1}{N-1}}\left(\frac{\gamma}{\rho}\right)^{\frac{2N-5}{3(N-1)}}\omega^{-\frac{7}{3}\left[1+\frac{3}{7(N-1)}\right]}{\rm \ ,}\\ 
        &S_{\eta}(k) =C_{KZ}^k \langle P \rangle^{\frac{1}{N-1}}\left(\frac{\gamma}{\rho}\right)^{-\frac{3}{2(N-1)}}k^{-3\left[1+\frac{1}{2(N-1)}\right]}{\rm \ ,}\\
    \end{split}
    \label{Seta_eq}
\end{equation}
where $\langle P \rangle$ is the mean cascading energy flux supposed constant. The nondimensional (Kolmogorov-Zakharov) constant of Eq.~(\ref{Seta_eq}) is denoted $C^{\omega}_{KZ}$ and $C^k_{KZ}$ with $C^k_{KZ}=\frac{3}{2}C^{\omega}_{KZ}$. Note that the capillary wave spectrum was derived exactly in 2D with $N=3$ \cite{ZakharovJAMTP1967} but not in 1D to our knowledge. To obtain analytically such spectrum, the kinetic equation could be derived following the weak turbulence assumptions and methods described in~\cite{ZakharovBook1992,NazarenkoBook2011,NewellARFM2011}. To do so, since the first nonlinear orders ($N=3$ and $N=4$) vanish in 1D for $\omega \sim k^{3/2}$, the leading nonlinear order is then $N=5$. The constant energy flux, out-of-equilibrium, stationary solution of this kinetic equation should provide the Kolmogorov-Zakharov wave spectrum corresponding to the energy cascade through scales. Note that the locality hypothesis (i.e., the convergence of the collision integral in the kinetic equation on the solution found) should be also fulfilled, otherwise corrections could arise~\cite{NazarenkoBook2011}. Such theoretical work is beyond the scope of our experimental study, and thus only dimensional predictions and numerical analysis are used here. Our predictions of Eq.\ (\ref{Seta_eq}) for $N=5$ yield to $S_\eta(k)\sim k^\alpha$ and $S_\eta(\omega)\sim\omega^\beta$ with $\alpha=-27/8$ and $\beta=-31/12$.

\section{Experimental setup}Experiments were performed in a rectangular transparent plexiglass canal of length $L=15$~cm and width $L_y=2$~cm (see Fig.~\ref{setup}). This canal is filled up to a depth $h=2$~cm. A shaker located at an extremity generates waves in a narrow random frequency bandwidth $f_0\pm\Delta f$ with $f_0=3.5$~Hz and $\Delta f=1.5$~Hz. This randomness triggers wave interactions needed for building wave turbulence, contrary to a monochromatic forcing, $\Delta f=0$ (see Supp. Mat. \cite{SuppMat}). The wave propagation is unidirectional along $x$ due to the geometry of the system ($L\gg L_y$) but since the wave energy is distributed in the 2D space ($x$ and $y$), we talk about quasi-1D waves. Wave elevation $\eta(t)$ is measured at a single point over time using a home-made capacitive wire gauge (10 $\mu$m vertical resolution and 2 kHz sampling frequency leading thus to a maximum observable frequency $f_m=1$~kHz) \cite{FalconPRL2007}. A typical example of $\eta(t)$ is plotted in the inset of Fig.~\ref{setup}, the forcing scales are clearly visible whereas higher scales are embedded within the observed nonlinearities. A space-and-time resolved wave-field measurement, $\eta(x,t)$, is reached using a lateral camera (Basler - 200 fps) filming a 9 cm side view with horizontal and vertical resolutions of 38~$\mu$m. The wave elevation is monitored for both measurements during $\mathcal{T}=15$ min.

\begin{figure}[t!]
    \centering
    \includegraphics[width=1\linewidth]{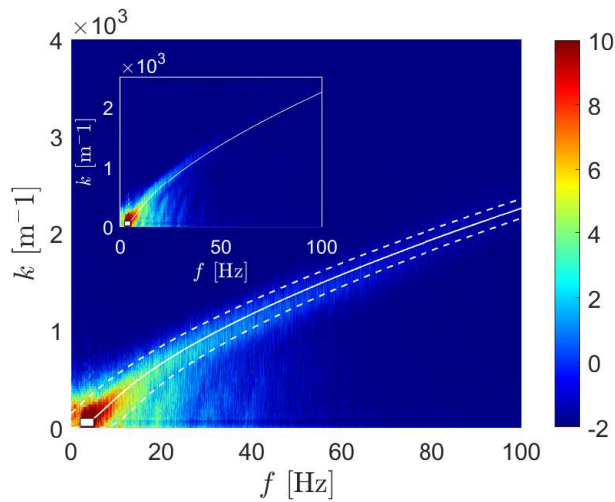} 
    \caption{Full spatio-temporal spectrum $S_{\eta}(k,\omega)$. Wave steepness $\epsilon=0.06$. Solid line: theoretical gravity-capillary dispersion relation $\omega(k)$. Dashed line: spread dispersion relation $\omega(k)\pm\delta_\omega$ with $\delta_\omega=40$~Hz. White rectangle: forcing range between 2 and 5 Hz. Inset: same for $\epsilon=0.03$. Log colorbar.}
    \label{spatio_temps_fig} 
\end{figure}

We perform experiments with mercury (density $\rho=13 500$~kg/m$^3$, surface tension $\gamma=400$~mN/m and kinematic viscosity $\nu=10^{-7}$~m$^2$/s). Its low viscosity is needed to minimize capillary wave dissipation to observe wave turbulence (using water instead leads to a cascade on a smaller inertial range). Its unwetting property also avoids the upward meniscus on the lateral wall that would hide the wave profile visualized by the camera.

To quantify the nonlinearity, we measure the wave steepness as $\epsilon=\left\langle \sqrt{\frac{1}{L}\int|\partial\eta/\partial x|^{^2}dx}\right\rangle$~\cite{BerhanuJFM2018}. Very weak steepnesses are used here, $\epsilon \leq 0.06$, to validate the WTT assumption of weak nonlinearity.

\section{Energy cascade}
From spatio-temporal measurements we perform space-time Fourier transforms to reach the full power spectrum $S_\eta(k,\omega)$ of wave elevations (see Fig.~\ref{spatio_temps_fig}). We observe that the energy injected at large scales (see white rectangle in Fig.~\ref{spatio_temps_fig}) is transferred to small scales following well the dispersion relation. In addition, the energy is found to be redistributed around the dispersion relation due to nonlinearities with a nonlinear spectral broadening $\delta_\omega$. This broadening is estimated for each $k$, by fitting the corresponding spectrum $S_\eta(k,\omega)$ by a Gaussian function of $\omega$. The standard deviation of this fit gives an estimate of $\delta_\omega$ whose average over the $k$ values is $\delta_\omega=40$~Hz (see dashed lines on Fig.~\ref{spatio_temps_fig}). Note that no coherent structures appear on the dispersion relation such as bound waves~\cite{HerbertPRL2010} or solitons~\cite{HassainiPRF2017}. 

\begin{figure}[t!]
    \centering
    \includegraphics[width=1\linewidth]{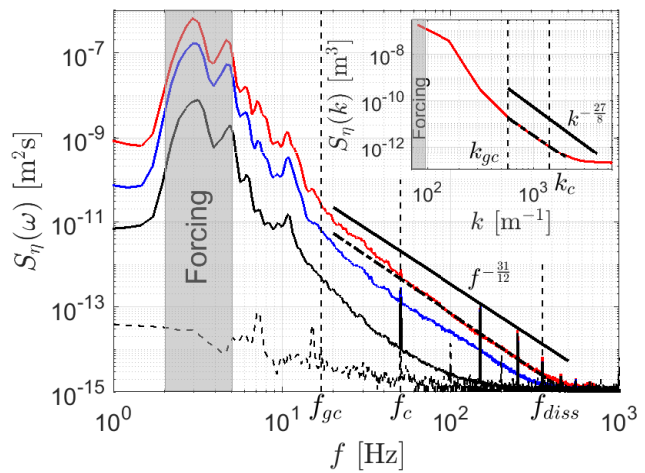} 
    \caption{Frequency spectra $S_{\eta}(f)$ for $\epsilon=0$, 0.006, 0.03, and 0.06 (from bottom to top). Bottom curve: experimental noise (no forcing). Grey area: forcing bandwidth. Dash-dotted line: $f^{-2.7}$ best fit. Solid line: predictions in $f^{-31/12}$ of Eq.~(\ref{Seta_eq}) with $N=5$. Inset: Wavenumber spectrum $S_{\eta}(k)$ for $\epsilon=0.06$. Dash-dotted line: $k^{-3.2}$ best fit. Solid line: predictions in $k^{-27/8}$ of Eq.~(\ref{Seta_eq}) with $N=5$.}
    \label{spectraOmeg}
\end{figure}

Figure~\ref{spectraOmeg} shows the frequency power spectrum $S_{\eta}(\omega)$ computed from the single-point measurement. The injected energy at large scales cascades first for $f<f_{gc}\simeq17$ Hz with visible forcing harmonics. For $f \gtrsim f_c$, we observe a frequency power-law cascade on a decade in the capillary range. We measure that the cascade fits in $f^{-2.7\pm0.2}$ which is found to be in good agreement with predictions in $f^{-31/12}$ of Eq.~(\ref{Seta_eq}) with $N=5$ (see solid line in Fig.\ \ref{spectraOmeg}). The exponent uncertainty is estimated by slightly changing the fitting frequency range. Note that predictions of Eq.~(\ref{Seta_eq}) with $N=3$ (respectively, $N=4$) leads to spectrum exponents in $f^{-17/6}$ and $k^{-15/4}$ (respectively, in $f^{-8/3}$ and $k^{-7/2}$) which could be also compatible with the experimental ones. However, no resonant wave interaction is involved theoretically in the capillary regime at these orders ($N=3$ and $N=4$). Note also that the experimental power law is slightly steeper ($f^{-3.3\pm 0.2}$) within $f\in [f_{gc}, f_{c}]$ Hz. Indeed, in this intermediate range, both capillarity and gravity contributions are important and deeply entangled so that no prediction exists either dimensionally or by WTT \cite{FalconARFM2022}, contrary to pure capillary regime ($f \gtrsim f_c$) where the gravity contribution to the dispersion relation is negligible ($<5\%$). Using the spatio-temporal measurement averaged on time, we plot the wavenumber spectrum $S_\eta(k)$ in the inset of Fig.~\ref{spectraOmeg}. The capillary cascade fits in $k^{-3.2\pm0.2}$ also close to the predictions in $k^{-27/8}$ of Eq.~(\ref{Seta_eq}) with $N=5$. Using $S_\eta(k)dk=S_\eta(\omega)d\omega$, $\omega(k)\sim k^{3/2}$, and assuming $S_\eta(k)\sim k^\alpha$, and $S_\eta(\omega)\sim \omega^\beta$, one has $2\alpha/[3\beta+1]=1$. The experimental ratio leads to $0.90\pm0.13$, showing thus consistency between the frequency and wavenumber spectra. The smaller inertial range for $S_{\eta}(k)$ compared to $S_{\eta}(\omega)$ is usual in wave turbulence \cite{BerhanuJFM2018,HerbertPRL2010}. It is linked, using the dispersion relation, to the lower frequency resolution of the camera than the wire gauge one.

\begin{figure}[t!]
    \centering
    \includegraphics[width=1\linewidth]{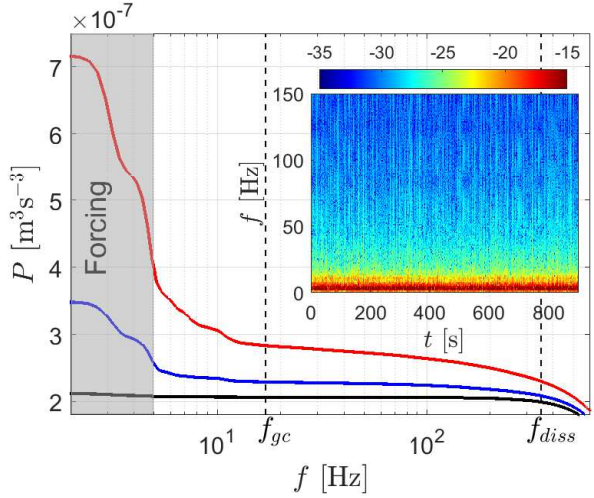} 
    \caption{Evolution of the energy flux $P$ with the frequency $f$ for $\epsilon=0.006$, 0.03, and 0.06 (from bottom to top). Grey area: forcing bandwidth. Averaged $P$ values (between $f_{gc}$ and $f_{diss}$): $\langle P \rangle=$2.04, 2.19 and $2.53\ 10^{-7}$~m$^3$s$^{-3}$. Inset: Time-frequency plot of $S_\eta(\omega,t)$ using a temporal step $\delta_t=0.50$~s and $\epsilon=0.06$. Log colorbar.}
    \label{P_f}
\end{figure}

We focus now on the energy flux cascading through the scales. It is estimated by $P(\omega^*)=\int_{f^*}^{f_{m}}E(\omega)\Gamma(\omega)d\omega$, using the total wave energy $E=gS_\eta(\omega)+(\gamma/\rho)k^2S_\eta(\omega)$ and the viscous dissipation rate of energy $\Gamma=k\sqrt{\nu\omega/2}$ for a contaminated interface~\cite{DeikePRE2014}. Figure~\ref{P_f} shows that the flux $P$ is almost constant in the inertial range [$f_{gc}$, $f_{diss}$] as expected theoretically. The temporal evolution of the Fourier modes $S_\eta(\omega,t)$ is plotted in the inset of Fig.~\ref{P_f}. The energy cascade is found to be stationary and continuous over the frequency scales (as also observed in Fig.~\ref{spectraOmeg}). No strong fluctuation occurs confirming the absence of coherent structures. The energy injected at the forcing scales thus cascades continuously over scales as expected by wave turbulence. Knowing the value of $\langle P \rangle=\int_{f_{c}}^{f_{diss}}P(\omega)d\omega$, we can thus infer experimentally the Kolmogorov-Zakharov constant $C_{KZ}^{\omega}$ using the spectrum fit in Fig.~\ref{spectraOmeg} and Eq. (\ref{Seta_eq}) with $N=5$. One finds $C_{KZ}^{k}=3C_{KZ}^{\omega}/2=(5.4 \pm 0.2)\ 10^{-3}$. Note that this constant is smaller than the ones measured and predicted for 2D capillary wave turbulence \cite{DeikePRE2014}. For instance, using Eq.~(\ref{Seta_eq}) with $N=3$ (2D) and $N=5$ (1D), one expects $C_{KZ}^{1D}\simeq C_{KZ}^{2D}/23$ for the same magnitude of the spectra, and typical values of $k=3\ 10^3$ m$^{-1}$ and $\langle P \rangle=10^{-7}$ m$^3$s$^{-3}$.

\section{Wave interactions}
We discuss now the role of nonlinear wave interactions. In a 1D pure capillary regime, no nontrivial \textit{resonant} interaction exists for $N=3$ or 4 (i.e., Eq.~(\ref{pureIntN4}) has no nontrivial solutions with $N=3$ or 4 with $\omega(k)=(\gamma/\rho)^{1/2}k^{3/2}$)~\cite{NazarenkoBook2011,KochurinJETP2020,ZakharovPR2004}. However, \textit{quasi-resonances} can exist at these low orders (i.e., Eq.~(\ref{quasiIntN4}) with $N=3$ or 4 has solutions) in a 1D pure capillary regime due to the nonlinear spreading of the dispersion relation (see Fig.~\ref{spatio_temps_fig})~\cite{KochurinJETP2020}. 

\begin{figure}[t!]
    \centering
    \includegraphics[width=1\linewidth]{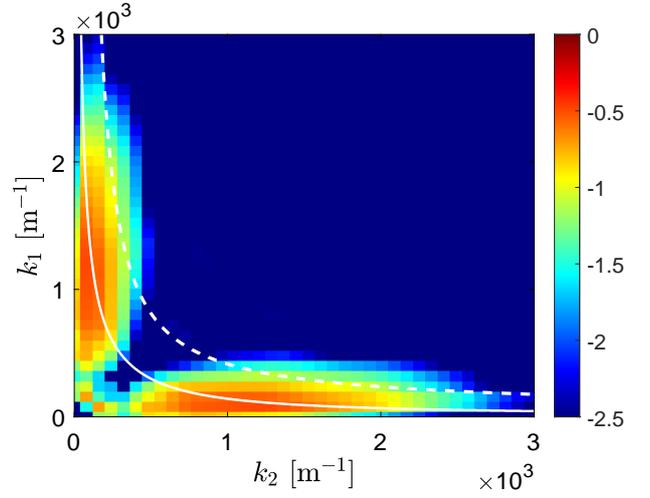}
       \caption{Bicoherence $B(k_1,k_2)$ for $\epsilon=0.06$. Solid line: \textit{resonant} interaction solutions of Eq.~(\ref{pureIntN4}) with $N=3$ and $\omega^2=gk+(\gamma/\rho)k^3$. Dashed line: borders of the \textit{quasi-resonant} interaction area, solutions of Eq.~(\ref{quasiIntN4}), $N=3$ with $\delta_\omega=40$~Hz. Log colorbar.}
    \label{Bico}
\end{figure}

First, we experimentally quantify three-wave interactions (i.e., $k_1+k_2=k_3$) by computing the normalized third-order correlation in $k$ of the wave elevations, called bicoherence~\cite{PunzmannPRL2009,AubourgPRF2016}
\begin{equation}
 B(k_1,k_2)=\frac{|\langle \eta_{k_1}^*\eta_{k_2}^*\eta_{k_1+k_2}\rangle|}{\sqrt{\langle| \eta_{k_1}\eta_{k_2}|^2\rangle\langle| \eta_{k_1+k_2}|^2\rangle}} {\rm \ ,}
\end{equation}
where $^*$ denotes the complex conjugate, the normalization being chosen to bound $B$ between 0 (no correlation) and 1 (perfect correlation). Figure~\ref{Bico} shows $B(k_1,k_2)$ for fixed $\epsilon=0.06$ (see Supp. Mat. \cite{SuppMat} for other values). The results are in agreement with the predictions for the resonance location [see solid line, solution of Eq.~(\ref{pureIntN4}) with $N=3$] and the quasi-resonance boundaries [see dashed lines, solutions of Eq.~(\ref{quasiIntN4}) with $N=3$]. These 1D solutions exist because of the coexistence of the gravity and capillary regimes in the dispersion relation. This probably explains why the frequency spectrum in Fig.\ \ref{spectraOmeg} is steeper in the range $f \in[f_{gc}$, $f_c$] than in the pure capillary range ($f>f_c$). Indeed, in this pure capillary regime, no resonant solution of Eq.~(\ref{pureIntN4}) exists with $N=3$ and $\omega(k)=(\gamma/\rho)^{1/2}k^{3/2}$. This is experimentally confirmed since the bicoherence is observed to vanish at large $k$. Thus, the three-wave resonant process occurring in 1D near the gravity-capillary crossover cannot generate wave turbulence in the pure 1D capillary regime ($f>f_c$).

\begin{figure}[t!]
    \centering
    \includegraphics[width=1\linewidth]{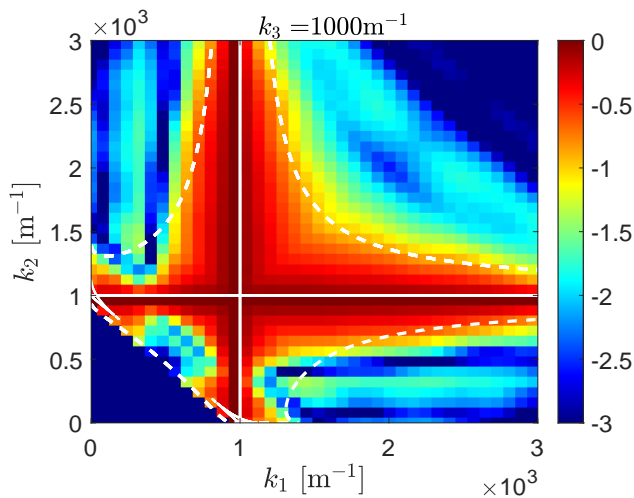}
       \caption{Tricoherence $T(k_1,k_2,k_3)$ for fixed $k_3=1000$ m$^{-1}$ and $\epsilon=0.06$ ($2\leftrightarrow2$). Solid lines: \textit{resonant} interaction solutions of Eq.~(\ref{pureIntN4}) with $N=4$ and $\omega^2=gk+(\gamma/\rho)k^3$. Dashed lines: borders of \textit{quasi-resonant} interaction area, solutions of Eq.~(\ref{quasiIntN4}), $N=4$ with $\delta_\omega=40$~Hz. Log colorbar.}
    \label{Trico}
\end{figure}

To highlight four-wave interactions (i.e., $k_1+k_2=k_3+k_4$), we compute the fourth-order correlation of the wave elevations (or tricoherence)~\cite{CampagnePRF2019}
\begin{equation}
T(k_1,k_2,k_3)=\frac{|\langle \eta_{k_1}^*\eta_{k_2}^*\eta_{k_3}\eta_{k_1+k_2-k_3}\rangle|}{\sqrt{\langle| \eta_{k_1}\eta_{k_2}|^2\rangle\langle|\eta_{k_3}\eta_{k_1+k_2-k_3}|^2\rangle}} {\rm \ .}
\end{equation}
Figure~\ref{Trico} shows $T(k_1,k_2,k_3)$ for fixed $k_3$ and $\epsilon=0.06$ (see Supp. Mat. \cite{SuppMat} for other values). It shows the occurrence of a large number of four-wave quasi-resonances within the area bounded by Eq.~(\ref{quasiIntN4}) with $N=4$ and $\delta \omega=40$ Hz (dashed lines). Note that four-wave pure resonances of Eq.~(\ref{pureIntN4}) with $N=4$, are also observed (see solid lines forming a cross centered on $k_1=k_2=k_3$). However, they do not lead to energy transfers between waves since $k_1=k_3$ and $k_2=k_4$, and are thus called degenerated or trivial resonances. Note also that other resonant solutions due to the gravity-capillary contribution appear close to $k_1=k_3$, $k_2=0$ and to $k_2=k_3$, $k_1=0$ (see small oblique solid lines). These nontrivial solutions can theoretically lead to a cascade but they vanish in the pure capillary regime ($f>f_c$ - see Supp. Mat. \cite{SuppMat}).

To sum up, we observe low-order resonant interactions ($N=3$ and 4) near the gravity-capillary crossover but they vanish in a pure capillary regime. Quasi-resonances are also present at these orders but, according to WTT, they are dynamically irrelevant to generate wave turbulence and are dominated by resonant interactions existing at the lowest order~\cite{NazarenkoBook2011}. It is thus necessary to consider higher-order exact resonances to understand the capillary cascade observed for $f>f_c$ in Fig.~\ref{spectraOmeg}. Equation~(\ref{pureIntN4}) has indeed nontrivial solutions for $N=5$ in a 1D pure capillary regime. This five-wave process is hence expected to be the dominant one generating wave turbulence for $f>f_c$.

To verify the existence of five-wave interactions, we compute the normalized fifth-order correlation in $k$ of the wave elevations, called quadricoherence. To our knowledge, this correlation has never been used before to analyze wave turbulence. Analogous to the above bi- and tricoherence used to explore lower-order correlations, we define quadricoherence as


\begin{equation}
 Q=\frac{|\langle \eta_{k_1}^*\eta_{k_2}^*\eta_{k_3}^*\eta_{k_4}\eta_{k_1+k_2+k_3-k_4}\rangle|}{\sqrt{\langle| \eta_{k_1}\eta_{k_2}\eta_{k_3}|^2\rangle\langle|\eta_{k_4}\eta_{k_1+k_2+k_3-k_4}|^2\rangle}} {\rm \ ,}
 \end{equation}
 
 \begin{figure}[t!]
    \centering
    \includegraphics[width=1\linewidth]{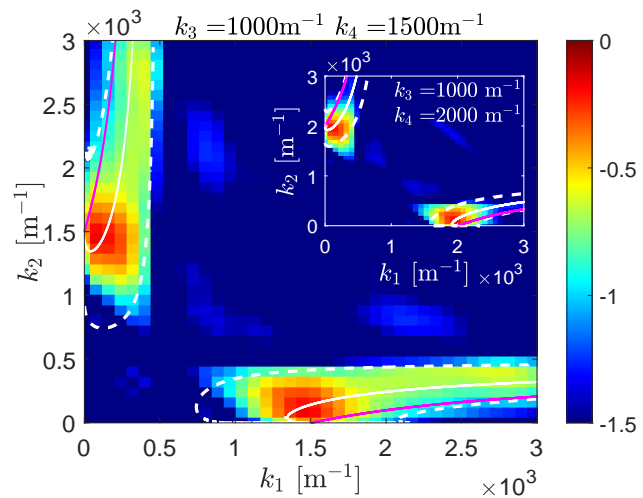} 
       \caption{Quadricoherence $Q(k_1,k_2,k_3,k_4)$ for fixed $k_3=1000$ m$^{-1}$, $k_4=1500$ m$^{-1}$ and $\epsilon=0.06$ ($3\leftrightarrow2$). Solid white lines: \textit{resonant} interaction solutions of Eq.~(\ref{pureIntN4}) with $N=5$ and $\omega^2=gk+(\gamma/\rho)k^3$. Dashed lines: borders of \textit{quasi-resonant} interaction area, solutions of Eq.~(\ref{quasiIntN4}), $N=5$ with $\delta_\omega=40$~Hz. Magenta solid lines: \textit{resonant} interaction solutions of Eq.~(\ref{pureIntN4}) with $N=5$ and $\omega^2=(\gamma/\rho)k^3$. Inset: same for $k_3=1000$ m$^{-1}$ and $k_4=2000$ m$^{-1}$. Log colorbar.}
    \label{Quadrico}
\end{figure}

Figure~\ref{Quadrico} shows $Q(k_1,k_2,k_3,k_4)$ for fixed $k_3$, $k_4$ and $\epsilon=0.06$ (see Supp. Mat. \cite{SuppMat} for other values). Quadricoherence shows the occurrence of a large number of five-wave resonances (solutions of Eq.~(\ref{pureIntN4}) with $N=5$ and $\omega^2=gk+(\gamma/\rho)k^3$, see white solid lines) and of quasi-resonances within the area bounded by Eq.~(\ref{quasiIntN4}) with $\delta \omega=40$ Hz (see dashed lines). Resonant solutions for $N=5$ in a pure capillary regime are also shown in Fig.~\ref{Quadrico} (magenta solid lines), by solving numerically Eq.~(\ref{pureIntN4}) with $N=5$ and $\omega^2=(\gamma/\rho)k^3$, i.e., solving
\begin{equation}
    k_1^{3/2}+k_2^{3/2}+k_3^{3/2}-k_4^{3/2}=(k_1+k_2+k_3-k_4)^{3/2}.
    \label{pureIntN5}
\end{equation}
The lower agreement of the magenta curve with data is due to the influence of the gravity regime on the smallest wavenumber of the pentad. The presence of five-wave resonances (see white solid line and green data) which subsists theoretically and experimentally in the pure capillary range (in contrast to $N=3$ or 4) indicates that the five-wave process is probably responsible for the capillary wave turbulence observed in 1D. Such 1D system thus strongly simplifies the problem since we only solved Eq.~(\ref{pureIntN4}) with $N=5$, i.e., Eq.~(\ref{pureIntN5}), to determine the location of the exact resonances in the spectral space. Note that the quadricoherence presented here refers to a $3\leftrightarrow2$ process (i.e., $k_1+k_2+k_3=k_4+k_5$), the $4\leftrightarrow1$ process (i.e., $k_1+k_2+k_3+k_4=k_5$) is not relevant here since present only in the gravity range. 

We now check that the main assumptions of WTT are well validated experimentally.



\begin{figure}[t!]
    \centering
    \includegraphics[width=1\linewidth]{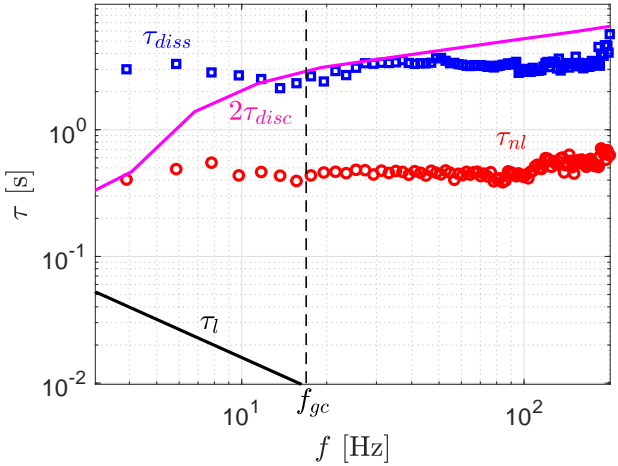} 
       \caption{Wave turbulence timescales versus the frequency scale. Solid line: linear timescale $1/\omega$. Nonlinear timescale $\tau_{nl}$ (circles) and dissipation time $\tau_{diss}$ (squares) estimated from wave turbulence decaying experiments. Magenta solid line: discreeteness time $\tau_{disc}$.}
    \label{timescales}
\end{figure}

\section{Timescales}
Weak turbulence theory assumes a timescale separation
\begin{equation}
    \tau_l(f)\ll\tau_{nl}(f)\ll[\tau_{diss}(f) {\rm \  and \ } \tau_{disc}(f)] {\rm \ ,}
    \label{ineq_taunl}
\end{equation}
between the linear propagation time $\tau_l=1/\omega$, the nonlinear interaction time $\tau_{nl}$, the dissipation time $\tau_{diss}$, and the discreeteness time $\tau_{disc}$. Satisfying Eq.~(\ref{ineq_taunl}) within the whole inertial range of scales $f$, requires that the nonlinear effects are slower than the linear ones and faster than dissipation and discreeteness ones, hence a nonlinear cascade of energy over scales is possible without finite size effect and dissipation. 

We use decay measurements to reach an estimation of these timescales. $\tau_{nl}$ (resp. $\tau_{diss}$) is inferred from the fast (resp. slow) decay of Fourier modes~\cite{CazaubielPRL2019} (see Supp. Mat. \cite{SuppMat}). $\tau_{disc}$ is computed as the number of eigenmodes found in a frequency band divided by this bandwidth, taking into account both transverse and lateral eigenmodes~\cite{FalconPRL2020}. No discreteness effect appears when $\tau_{nl}<2\tau_{disc}$ (i.e., nonlinear spectral widening $>$ half frequency separation between adjacent eigenmodes).

Figure~\ref{timescales} shows clearly that the timescale separation of Eq.~(\ref{ineq_taunl}) is fulfilled regardless of $f$. Moreover, $\tau_{diss}$ and $\tau_{nl}$ are found to be roughly independent of the scale $f$, due to finite size effects of the system. Indeed, $\tau_{diss}$ is the same order of magnitude as the linear viscous dissipation by surface boundary layer of the main lateral eigenmode $\tau_{diss}=2\sqrt{2}/[k_{L_y/2}\sqrt{\nu\omega(k_{L_y/2})}] \approx3$~s \cite{Lamb1932,DeikePRE2012}. The constant nonlinear time $\tau_{nl}\approx0.4$~s is probably due to cumulative energy transfer from this eigenmode in addition to the usual contribution by nonlinear wave interactions \cite{CazaubielPRL2019}. To sum up, the timescale separation assumed by WTT is verified although finite size effects are present (leading to constant values for $\tau_{nl}$ and $\tau_{diss}$) but not enough ($\tau_{nl}\ll \tau_{disc}$) to prevent the occurrence of wave turbulence. 

\section{Conclusion}
Small-scale wave turbulence had never been demonstrated experimentally in a quasi-1D system until now. Here, we evidenced the first experimental observation of quasi-1D capillary-wave turbulence. We show that five-wave resonant interactions are the lowest-order resonant process involved in the pure capillary regime, and is probably the one leading to the observed 1D capillary-wave turbulence. Experimental wave spectra are found compatible with the corresponding dimensional analysis prediction. The main WTT assumptions are also verified (weak nonlinearity, timescale separation, and constant energy flux). 
As done in 1D gravity waves~\cite{dyachenko1995five}, a theoretical confirmation by WTT of this five-wave process in 1D capillary wave turbulence would be of primary interest. Beyond fluid mechanics, our work could pave the way to other studies in different wave turbulence fields (such as ferrohydrodynamics~\cite{BoyerPRL2008}, quantum fluids~\cite{LvovJETP2010}, elasticity~\cite{DuringPRL2006} or hydro-elasticity~\cite{DeikeJFM2013}), due to easier calculations, simulations and measurements in such 1D geometry.

\acknowledgments
We thank E. Kochurin and S. Nazarenko for fruitful discussions. This work was supported by the French National Research Agency (ANR DYSTURB project No. ANR-17-CE30-0004), and by the Simons Foundation MPS N$^{\rm o}$651463-Wave Turbulence.


\end{document}